# A comparison of time-dependent Cloudy astrophysical code simulations with experimental X-ray spectra from keV laser-generated argon plasmas

N. Rathee[1], F.P. Keenan[1], R.J.R. Williams[2], G.J. Ferland[3], S.J. Rose[4], S. White[1] and D. Riley[1]

[1] School of Mathematics and Physics, Queen's University Belfast, University Road, Belfast BT7 1NN, UK
[2] School of Engineering Mathematics and Technology, University of Bristol, Bristol BS8 1TW, UK
[3] Department of Physics and Astronomy, University of Kentucky, Lexington, KY 40506
[4] Plasma Physics Group, Imperial College London, South Kensington Campus, London SW7 2AZ, UK

## Abstract

We have generated strongly photoionized Ar plasmas in experiments designed to use primarily X-ray L-shell line emission generated from Ag foils irradiated by the VULCAN high-power laser at the UK Central Laser Facility. The principle of the experiment is that use of line emission rather than the usual sub-keV quasi-blackbody source allows keV radiation to play a more dominant role compared to softer X-rays and thus mimic the effect of a blackbody with a higher effective spectral temperature. Our aim is to reproduce in the laboratory the extreme photoionization conditions found in accretion-powered astrophysical sources. In this paper, we compare the experimental results on K-β X-ray Ar spectra with modelling using the time-dependent version of the Cloudy astrophysical code. The results indicate that photoionized laboratory plasmas can be successfully modelled with codes such as Cloudy that have been developed for application to astrophysical sources. Our comparison of simulation and experiment shows that the flux of sub-keV photons that photoionize the outer-shell electrons can have a significant effect, and that detailed measurements of the X-ray drive spectrum across all photon energy ranges are crucial for accurate modelling of experiments.

## 1. Introduction

Photoionization-dominated plasmas are found in accretion-powered astrophysical objects [1-7], where heating is driven primarily by radiation processes rather than collisional ones. The degree to which photoionization dominates in a steady-state plasma is defined by the photoionization parameter, $\xi = 4\pi F/n_e$, where $F$ is the incident flux and $n_e$ the electron density. Typical values of $\xi$ for different astrophysical environments range from 1 erg cm s$^{-1}$ for moderately ionized to 1000 erg cm s$^{-1}$ [7] for highly ionized gas. Several attempts have been made to create such plasmas in the laboratory to benchmark photoionization codes used to model astrophysical sources, such as Cloudy [8], [9]. Foord et al. [10] generated their laboratory plasmas by using X-rays from the Z-pinch device at the Sandia National Laboratory to heat and decompress a solid target. The resulting photoionized plasma achieved $\xi \sim 20 - 25$ erg cm s$^{-1}$. A reasonable agreement in the electron temperature and ionization state of Fe with Cloudy predictions was achieved by treating Fe resonance lines as optically thick. Subsequently, Fujioka et al.[11] used the high-power GEKKO-XII laser facility, achieving $\xi = 6$ erg cm s$^{-1}$ and a radiation temperature of 500 eV for 0.2 ns. More recently, several further experiments have been performed at the Z facility by Loisel et al.[12], Mancini et al.[13] and Mayes et al. [14], achieving values of up to $\xi = 60$ erg cm s$^{-1}$. We note that the photoionization parameter strictly only applies to steady-state plasmas, which have not been truly achieved to date in the laboratory. Nevertheless, estimates of $\xi$ are useful in assessing the degree and importance of photoionization in the laboratory plasma compared to astrophysical conditions.

White et al.[15] used the VULCAN laser at the UK Central Laser Facility, achieving $\xi = 45$ erg cm s$^{-1}$ although once again not in steady-state. In this experiment, a novel technique developed by us [16] was successfully implemented, where 3–4 keV L-shell radiation from a Sn foil was used to ionize an Ar target, rather than the quasi-blackbody source normally employed. The line source provides not only a high X-ray flux but also means that the keV radiation dominates over softer X-rays. This mimics the effect of a higher spectral temperature on the photoionization of Ar, with the relative importance of inner-shell photoionization increased relative to outer-shell, as expected in astrophysical plasmas with a high radiation temperature. The steady-state version of Cloudy used to model the VULCAN data predicted a significantly higher Ar ionization state than seen experimentally. Primary causes for this are believed to be the importance of the nanosecond timescale of the experiment and also the strong gradient in flux within the sample plasma.

In a follow-up VULCAN experiment, Riley et al. [17] achieved $\xi > 50$ erg cm s$^{-1}$ when Ar gas was photoionized using 3-4 keV line radiation, but this time from a double-sided Ag foil irradiated on both sides by the laser, which reduced the flux gradient significantly. Using an in-house code [17], the collisional ionisation rate for Ar+ was estimated to be ~$10^{10}$ s$^{-1}$, giving a characteristic time for ionisation of ~100 ps. This is shorter than the 1 ns pulse duration but comparable to the rise time of the X-ray flux, and early on it is several hundred picoseconds. The recombination rates are also much slower, and the higher ionisation stages have ionisation times that are longer, which justifies the need to use the time-dependent Cloudy code.

A comparison of the Ar K-$\beta$ spectrum of an experimental shot for 103 mbar gas fill with a simulation from the time-dependent version of Cloudy presented in [17] revealed that our Cloudy results overestimated the intensities of emission lines from more highly ionized Ar charge states. This was due to the Cloudy simulations only being for the central region of the Ar plasma, while the experimental data were spatially integrated over the whole emission region. In this paper, we compare experimental data to a modified Cloudy simulation model accounting for the profile of incident intensities along the radial geometry of the target and discuss the influence of L-shell and broadband fluxes on the ionization state of the Ar gas target. The paper is organized as follows. In section 2, we briefly describe the experimental setup and data shots, while section 3 presents the parameters used to model the photoionized plasmas with Cloudy. Results are given in section 4, and finally in section 5 we discuss our findings and provide a brief description of our upcoming experiments which are informed by these.

## 2. Experimental setup

The experimental data employed in the present paper were obtained using the VULCAN laser at the UK Central Laser Facility. Here we provide a brief overview of the experimental setup and primary diagnostics used, with a more detailed description given by Riley et al. [17]. Figure 1 shows the target structure and laser irradiation configuration. The experiment consisted of an octagonal gas-cell target with Ag-coated CH foils at each end of thickness $3.8 \, \mu m$ and $18.6 \, \mu m$. Cells were prefilled with argon gas at pressures ranging from 10 to 500 mbar, while the Ag coatings were 467 nm thick, and separated by either 2 mm or 3 mm. The Ag foils were irradiated with 3 laser beams on each side, delivering up to 460 J per side of 527 nm wavelength laser energy, in a ~1.5 ns duration pulse. These beams were focused to a focal spot of approximately 300 µm diameter on the foil target, resulting in typical peak irradiances of about 4×10$^{14}$ Wcm$^{-2}$. Our previous experiments on VULCAN indicate that approximately 1% of the laser energy is emitted from the foil into the gas-fill, in the form of L-shell X-ray emission in the 3–4 keV range [18].

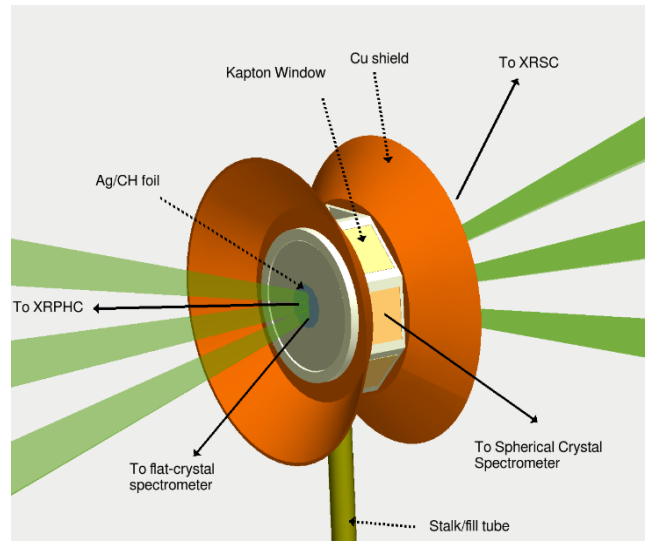

*Figure 1 Visrad figure of the gas cell target. The gas cell and the laser irradiation are symmetric with three beams focused on each side as described in the text. There were X-ray pinhole cameras (XRPHC) and flat crystal spectrometers viewing both Ag/CH foils. The X-ray streak camera (XRSC) only viewed one side. A second spherical crystal spectrometer viewed the plasma from the opposite side to the one shown. As discussed in the text, the spherical crystal spectrometers spatially resolved in mutually orthogonal directions.*

Hydrodynamic simulation using the Hyades code [19], indicated that a shock of about 30 Mbar is launched into the CH substrate that results in a transmitted shock of about 300 Kbar in the Ar gas, travelling at a speed of approximately 200 km/s. Since the Ag/CH foils are recessed by 0.5 mm to prevent direct viewing of the source by our diagnostics (see below), the shock would emerge into view after 2.5 ns, by which time the X-ray flux, which drives our signal, is over. Earlier test shots with Al/CH foils, which have no line emission above the 3.2 keV threshold for K-shell ionisation, indicated no significant signal, confirming that photoionisation is the driver of our signal.

The principal diagnostics for the experiment were a pair of spherical Bragg crystal spectrometers. One of these was fitted with a mica crystal of 150 mm radius-of-curvature, used in 4th order to operate in the Ar K-α to K-β spectral range (~2950–3200 eV). This was oriented to give spatial resolution in the axial direction, which is that normal to the laser-irradiated foils. The other diagnostic was a quartz (11–20) crystal, also with 150 mm radius-of-curvature, oriented to give spatial resolution in the radial direction, defined as that parallel to the laser-irradiated foils.

The L-shell emission from the Ag foils was monitored using Bragg crystal spectrometers. We expect this emission to be accompanied by a broad continuum due to recombination radiation, unresolved transition arrays and bremsstrahlung from the high-Z laser plasma. Since the broadband flux was not monitored experimentally, for the simulations in Riley et al [17] we assumed a Planckian distribution modified by transmission through a CH substrate layer. We initially assumed an effective Planckian at ~170 eV, based on limited measurements in the ~1.5 keV region from an earlier experiment, and also assumed the cold opacity for the CH substrate that X-rays passed through. In addition, in [17] we saw that a better fit to the average ionisation was gained by reducing the effective radiation temperature to 142 eV (halving the assumed Planckian flux). As discussed below, in the present work we explore the effects on the simulations of using a more realistic X-ray spectrum and a warm plasma opacity for the CH layer.

In figure 2 we show the results of HYADES [19] simulations, where 500 ps into the simulation the peak radiation temperature is between $T_R$ = 140 – 150 eV. There is also heating of the CH such that the electron temperature is ~ 20 eV for most of the foil. The foil density and temperature as well as the radiation temperature will clearly be functions of time as the foil is driven by the laser. It will initially compress due to the shock and then decompress as it heats, although the $\rho R$ of the foil will remain at the initial value. As discussed in the next section, in the Cloudy simulations, we have adopted single values for the foil parameters and radiation temperature.

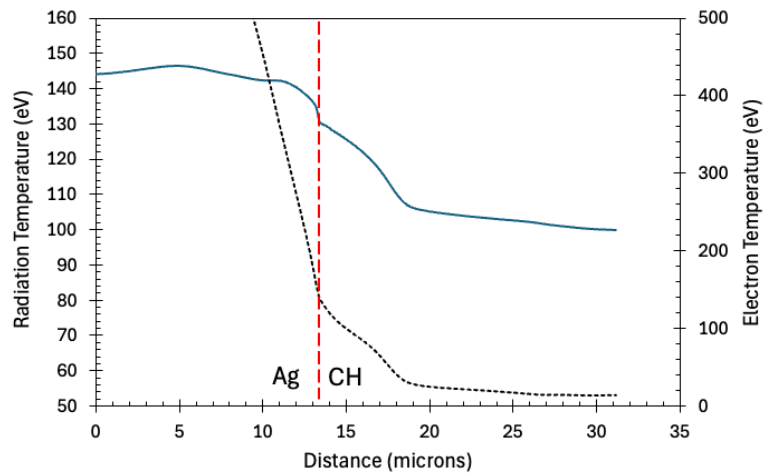

Figure 2: Hyades simulation of the Ag/CH foil driven by a square pulse laser of width 1.2 ns and a rise and fall time of 0.2 ns, showing radiation temperature (solid line) and electron temperature (dotted line) 500 ps into the simulation. The vertical line marks the boundary of the Ag and CH components of the foil. Peak intensity is $4 \times 10^{14}$ $Wcm^{-2}$. The Ag foil is 467 nm thickness, and the CH is 18.6 $\mu m$.

For some of the experimental shots we have K-$\beta$ emission line data that spectrally separates the emission from different ion stages, allowing a clear measurement of an averaged, effective ionisation that can be compared to simulation. In the other shots, a silicon nitride ($Si_3N_4$) slit was added to restrict the plasma size viewed, and for these the K-$\beta$ emission was too weak to be used. However, the higher spectral resolution obtained allowed a better comparison of the K-$\alpha$ emission with Cloudy predictions.

For the K-$\beta$ data we have compared Cloudy simulations with the first three data shots listed in table 1, where the gas pressure of Ar and the distance between foils are varied. The differences in foil distance affect the L-shell flux at the centre of the gas cell, which is listed for each case and was monitored in the experiment. We then considered the effect of changing the CH filter to a much thinner layer. This is expected to drive ionisation beyond the first few ion stages, and hence the K-$\beta$ emission would become much weaker as there as few electrons in the 3p orbital. In these cases, we investigate the K-$\alpha$ radiation from the shots fitted with the $Si_3N_4$ slit which, as noted already, have improved spectral resolution at the cost of a weaker overall signal.

Table 1: Data shots from the VULCAN experiment used in this paper, which all have within 6% of an average of 907 J of laser energy on target. The error bar in the pressure was typically +/- 4 mbar.

| Shot # | P (mbar) | CH thickness ($\mu m$) | Ag foil separation (mm) | L-shell flux (erg cm$^{-2}$s$^{-1}$) | Note |
|---|---|---|---|---|---|
| 75 | 55 | 18.6 | 3 | $1.4 \times 10^{18}$ | |
| 80 | 106 | 18.6 | 3 | $1.4 \times 10^{18}$ | |
| 93 | 55 | 18.6 | 2 | $3.15 \times 10^{18}$ | |
| 85 | 106 | 3.8 | 3 | $1.2 \times 10^{18}$ | Si$_3$N$_4$ slit K-$\alpha$ only |
| 86 | 106 | 18.6 | 3 | $1.4 \times 10^{18}$ | Si$_3$N$_4$ slit K-$\alpha$ only |

## 3. Cloudy Simulations

Our plasma simulations have been generated using the time resolving capability ([20], [21]) of the astrophysical photoionization code Cloudy 23.01([8], [9]). The output of the Cloudy simulations is integrated over a 2 ns period to improve consistency with the time-integrated experimental data: this time window encompasses the total duration of the driving X-ray pulse. For the quartz spectrometer used here, the data are also averaged over the axial direction, which is perpendicular to that of the spatial resolution for this crystal.

In the simulations, the sample gas is set at a nominal distance of $1\ cm$ from the X-ray source with a thickness of 0.1 cm, which represents the spatial integration of the lineout taken in the experimental data. As Cloudy requires that elemental abundances are input relative to the H abundance, the Ar densities were scaled relative to a (negligible) density of $1\ cm^{-3}$ for H. A constant 298 K blackbody is included to ensure that the gas is initially neutral.

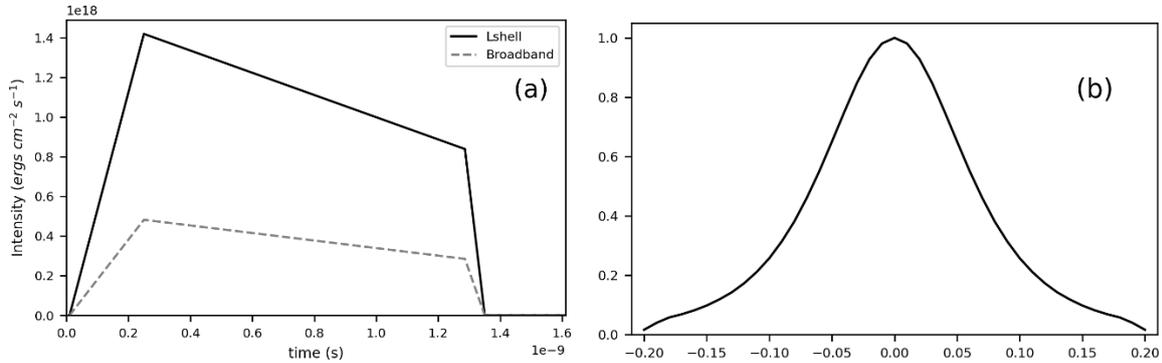

Figure 3: (a) Temporal profile of the incident X-ray fluxes used in the Cloudy. This is based on the measured L-shell history. A similar temporal profile but with different intensities is used for the other shot simulations. (b) Relative profile of the total incident flux along the radial direction at the centre of the gas cell (i.e. 1.5 mm from each foil).

In addition, two spectral energy distributions (SEDs) in the range 3.2-3.7 keV (Ag L-shell drive) and 0-2.5 keV (broadband), whose total flux is varying with time, are included. Total fluxes are provided at 4 points in time, and Cloudy interpolates the intermediate values using a linear equation. The timestep used in the simulation is 0.01 ns. Fluxes corresponding to these SEDs are negligible ($\sim \times 10^{-11}$ erg cm$^{-2}$s$^{-1}$) at the initial time and reach a maximum at ~0.2 ns. Figure 3(a) shows the temporal history of the L-shell drive and broadband intensities for Shot 75, assuming the broadband to be Planckian. The absolute peak values are shot dependent, but the temporal shape is the same for all shots.

In figure 3(b) we plot the normalised radial profile of the flux at the centre of a cell with a 3 mm separation between foils. To capture the effect of changes in the incident flux along the radial direction, we ran simulations 20 times for different incident intensities, following the relative scaling shown in Figure 3(b). The final time-integrated spectra are Abel transformed over the full extent of the radial profile, i.e. from –2 to +2 mm.

The L-shell spectrum used for input to Cloudy, based on detailed experimental measurements [18], is shown in Figure 4(a). For the Cloudy simulations, we divide it into 80 photon groups, equally spaced in energy. For the broadband SED we have taken two approaches. In the first, we have used the radiation temperature at the peak of the laser from the HYADES simulations and assumed a Planckian spectrum that is modified by passage through the CH foil whose opacity is calculated using the Ionised Material Package (IMP) [22] for a typical set of conditions at that time. In the case of the thicker CH foil (18.6 μm) we have taken $T_R$ = 146 eV, CH density = 1 gcm$^{-3}$ and CH $T_e$ = 20 eV. For the thinner CH foil (3.8 μm), the HYADES simulation gives a lower radiation temperature, but hotter CH and we adopt $T_R$ = 138 eV with density = 0.2 gcm$^{-3}$ and $T_e$ = 70 eV for the CH foil.

In the second approach, we have used the FLYCHK code [23] to generate typical spectra in the 0-2.5 keV regime for Ag. We divided the Ag foil in the HYADES simulation into 4 regions of differing temperature and density and the CH foil into 3 regions. The final spectrum is a superposition of these 7, with the assumption that there is no radiation coupling between these regions. In making comparisons between the two approaches, we have scaled the FLYCHK results to give the same total flux as the Planckian spectrum, prior to passing through the CH foil. In this way, in both cases the flux is consistent with the effective radiation temperature given by HYADES.

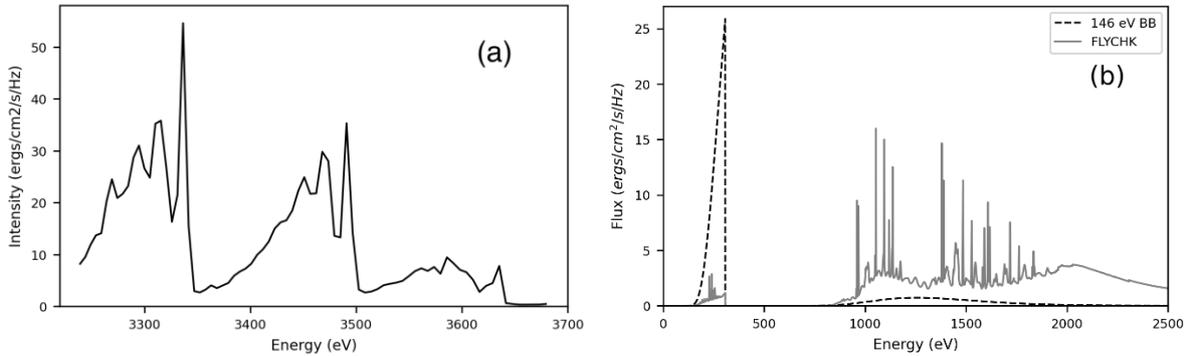

*Figure 4: Input SEDs for Cloudy simulations. (a) Normalised L-shell flux from the Ag foil, divided into 80 equally spaced photon groups. (b) Broadband flux corresponding to a Planckian with 146 eV spectral temperature (dashed line) and FLYCHK generated (solid line) modified to include the effect of the CH filters of thickness 18.6 μm that are radiatively heated to 20 eV.*

Figure 4(b) plots the broadband emission assuming a 146 eV Planckian passing through the CH filter at $T_e$ = 20 eV and $\rho$ = 1 gcm$^{-3}$. We compare it with the expected flux from FLYCHK. The expected flux at the Ar gas in the middle of a 3 mm gas cell is 4.8x 10$^{17}$ ergcm$^{-2}$s$^{-1}$ for the Planckian case and 1.1x10$^{18}$ ergcm$^{-2}$s$^{-1}$ for the detailed FLYCHK spectrum. The enhanced flux in the latter case is due to the photon distribution having significant weight in the ~1 keV region, well above the K-edge of carbon, thus reducing the effective opacity of the CH layer. As we see in the next section, the additional flux and the redistribution of the photon flux compared to the black-body case are both factors in the simulated fluorescence spectra for Ar.

## 4. Results

Figure 5 shows a comparison of the Ar K-β line emission measured by the Quartz crystal spectrometer with the Cloudy simulations for gas pressures of 55 and 106 mbar. The experimental results are for shots with 18.6 μm CH filter thickness and 3 mm separation between the Ag foils. For our simulations we use linewidths determined from fitting to the experimental data and line energies calculated with a Hartree-Fock atomic model as described in [17]. The linewidths vary, due it is believed to the excited states of the different ions stages that lead to satellite K-β features, which are slightly shifted from the principal lines but not by enough to be resolved. These parameters are used in conjunction with the spectral line intensities determined by the Cloudy model to generate the simulated spectra.

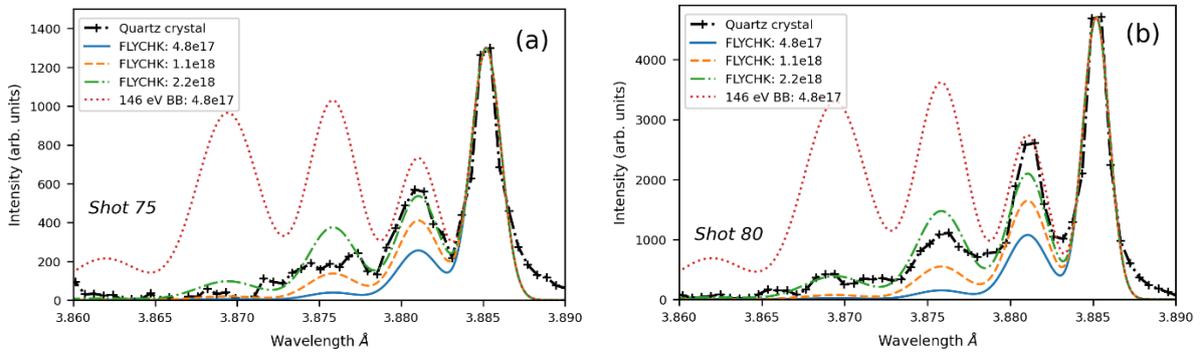

Figure 5: Comparison of the experimental Ar K-$\beta$ emission spectra with time-dependent Cloudy simulations for (a) 55 mbar and (b) 106 mbar gas pressures. For these shots, the distance between the Ag foils is 3 mm and the CH filter thickness is 18.6 μm. The L-shell incident flux intensity is $1.4 \times 10^{18}$ erg-cm$^{-2}$-s$^{-1}$ at the gas-cell centre. The line labels represent the broadband SED used in the simulation and their corresponding total flux (see text).

As we can see from figure 5, our use of the Planckian SED, with radiation temperature 146 eV and a warm opacity in the simulation, leads to a significant overestimation of the Ar gas ionisation (dotted line). This is not surprising because, as from figure 4(b) there is a significant flux expected in the spectral region close to the outer shell photoionization edges for Ar. The use of the FLYCHK SED with the higher expected transmission through the CH layer (flux on target ~$1.1 \times 10^{18}$ erg-cm$^{-2}$-s$^{-1}$), leads to theoretical spectra closer to experiment (orange dashed line) but still predicts a lower ionisation than experimentally observed. We have also included two other simulations for comparison. The first uses the same incident flux as the Planckian case and we can see that this significantly underestimates the ionization. This arises because the flux is now more heavily weighted towards ~1 keV photons away from the outer-shell ionisation potentials although not energetic enough for K-shell ionisation. We have also included a simulation where we double the expected flux for the FLYCHK SED. In these cases, we predict ionisation close to the experiment.

In figure 6 we show the results of K-α emission spectral simulations for shots 85 and 86, where a slit was added to the gas cell to improve resolution. In these cases. For the thinner foil case in figure 6(a), the use of a Planckian SED with 138 eV radiation temperature leads to a theoretical spectrum close to the experiments (red dotted line). We also see again that using

FLYCHK with a flux based on the same radiation temperature but with a higher transmission through the CH layer leads to a better (but still under ionised) spectrum whilst, as before, doubling the flux gets us closer to experiment.

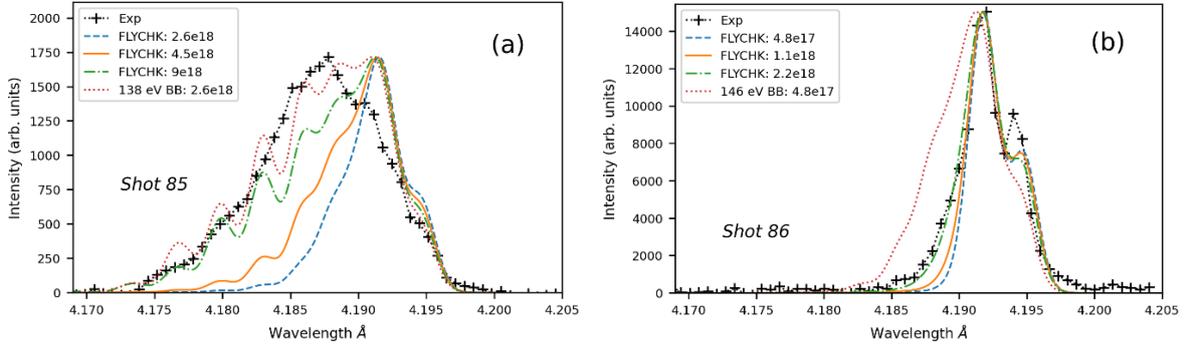

Figure 6: Comparison of the experimental Ar K-$\alpha$ emission spectra with time-dependent Cloudy simulations for gas cells with 106 mbar gas fill and 3 mm foil separation. (a) 3.8 $\mu m$ CH substrate, L-shell flux at cell centre is $1.2\times10^{18}$ erg-cm$^{-2}$s$^{-1}$. The line labels represent the broadband SED used in the simulation and their corresponding total flux. (b) 18.6 $\mu m$ CH substrate with L-shell flux $1.4\times10^{18}$ erg-cm$^{-2}$s$^{-1}$. The line labels represent the broadband SED used in the simulation and their corresponding total flux

This result is seen also in figure 6(b) for a thicker foil, where the Planckian case overestimates ionisation as it does for the K-$\beta$ data in figure 5. Once again, a higher than expected flux for the FLYCHK SED is needed to get a good matching to experiment.

In the data presented so far, we see a consistent picture that, when coupled to a realistic opacity for the CH foil, a quasi-Planckian predicts significantly higher ionisation than experiment. However, the use of a more realistic spectrum generated by FLYCHK requires a doubling of the expected flux to match the experimental data. We consider two potential reasons for this. One possibility is that the hydro-code underestimates the radiation temperature. An increase in 20% for the effective radiation temperature, $T_{Rad}$, would lead to a doubling of the flux, and a $T_{Rad} \sim 175$ eV. Note that this is close to the effective temperature initially considered in [17] which was found to lead to too high an ionisation when applied to a Planckian with warm CH opacity, whilst the lower effective temperature used here is consistent with HYADES employing multi-group opacities in 50 groups logarithmically spaced from 0.03 to 20 keV. We should also consider that the data are time-integrated and there will be an evolution of the opacity of the CH foil. Although we have adopted 20 eV as the characteristic temperature close to the peak of the X-ray pulse, in fact it varies from cold to over 200 eV average near the end of the X-ray flux. This would lead to a temporally varying input SED that we cannot as yet implement in Cloudy.

In figure 7 we show simulations for shot 93 where the foils are closer together (2 mm separation), resulting in a higher flux of $3.15 \times 10^{18}$ erg cm$^{-2}$ s$^{-1}$ for the L-shell. The CH foil thickness is 18.6 $\mu m$ and the gas pressure 55 mbar. For this case, a broad agreement between theory and experiment can also be achieved by using the FLYCHK SED, this time using the flux determined from the expected effective radiation temperature and transmission of the CH foil, without the need for arbitrary doubling of the flux. This conclusion is a little different from the data for 3 mm gas cells presented earlier. It is not clear why this should be, given that the evolution of the Ag plasma and CH foils should, in principle be similar to the data shown in Figure 5. One issue for consideration is whether the X-ray flux from opposite foils may have a stronger effect on the opacity of the foil opposite, and hence alter the opacity of the CH. Additional heating should lower the effective opacity and thus allow more of the flux through.

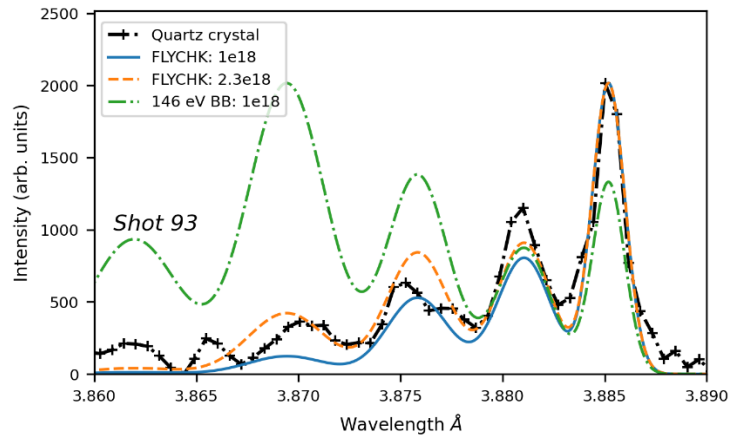

Figure 7: Comparison of the experimental Ar K-β emission spectrum with time-dependent Cloudy simulations for a gas cell of 55 mbar pressure with 18.6 μm CH substrate and 2 mm separation between foils. The Planckian SED significantly overestimates ionisation, while the FLYCHK SED predicts a spectrum closer to the experiment.

In figure 8 we show the K-β emission profiles for shot 93, generated using Cloudy, spatially resolved in the radial direction across the centre of the gas cell. Note that the notation $Ar^{n+}$ refers to the ionisation state of the Ar ion immediately after inner shell-ionisation. In both cases of the FLYCHK (a) and Planckian SEDs (b), we see that fluorescence emission from the lower ion stages comes from a wider region, as would be expected, for a flux profile as shown in figure 3(b). The effect of a Planckian SED is evident by the fact that in figure 8(b) the dominant ion stage is $Ar^{4+}$, whilst for the FLYCHK SED in figure 8(a) it is $Ar^{1+}$. This difference is driven by the additional flux at lower photon energies for the Planckian SED (see figure 4(b)), which allows for stronger photoionization of the 3s and 3p electrons.

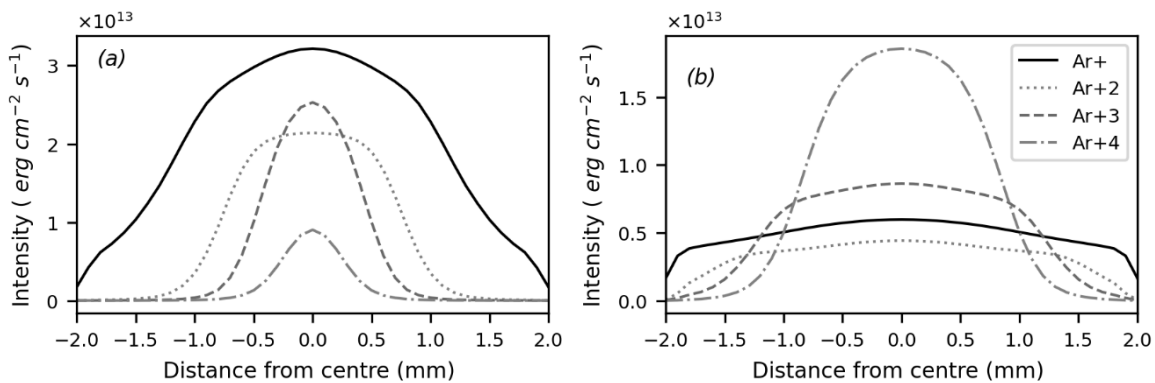

Figure 8: Time-integrated K-β line intensities from different Ar ionization states along the radial profile of the gas cell, calculated with Cloudy for shot 93. The gas pressure for this shot is 55 mbar, CH foil thickness 18.6 μm and Ag foil separation 2 mm, with L-shell flux of $3.15 \times 10^{18}$ erg-$cm^{-2}s^{-1}$ We assume a total flux of $1 \times 10^{18}$ erg-$cm^{-2}s^{-1}$ for the broadband drive, with (a) FLYCHK SED; (b) Planckian SED with 146 eV radiation temperature.

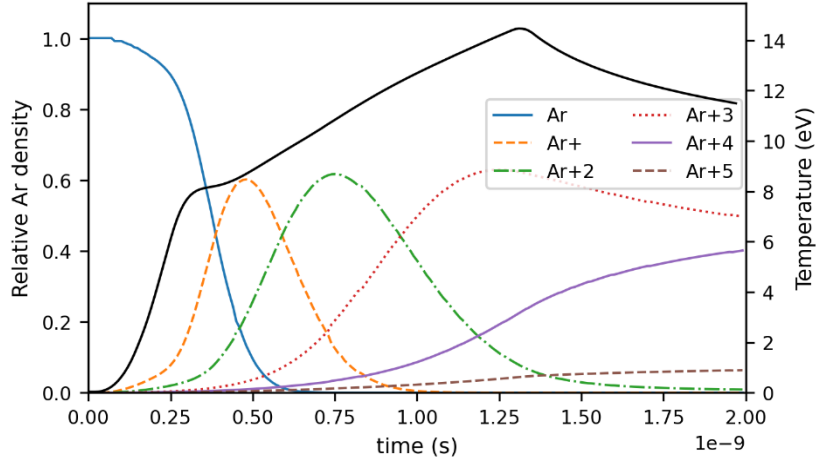

*Figure 9: Time-dependence of ion stage populations and electron temperature calculated by Cloudy at the centre of gas cell for shot 93 assuming a FLYCHK SED with total flux of $2.35 \times 10^{18}$ erg-cm$^{-2}$s$^{-1}$.*

Finally, in figure 9 we show the temporal history of the Ar ion stages, and the electron temperature predicted at the centre of the gas cell, for shot 93 assuming a FLYCHK SED with total flux of $2.35 \times 10^{18}$ erg-cm$^{-2}$s$^{-1}$. We see that the temperature continues to rise until the end of the X-ray flux at around 1.3 ns when the electrons are at ~14 eV. This figure clearly illustrates the dependence of our spectra on temporal integration, as the lower ion stages dominate emission early on. The higher ion stages rise in population as time progresses and a changing opacity, as discussed above, might have an influence on the relative intensity of the fluorescence from these stages. A time-gated spectrometer with even a moderate resolution of ~ 0.5 ns would show the dominance of the few times ionised stages in the middle of the X-ray drive pulse. However, a time-gated instrument would likely come at the cost of spectral resolution due to the more limited spatial resolution usually achieved in comparison to a direct X-ray CCD detector.

## 5. Discussion and future work

We have presented a comparison of laser-generated photoionized plasma experimental data with time-dependent Cloudy astrophysical code simulations, with the code able to reproduce experimental Ar K-$\alpha$ and K-$\beta$ emission line spectra reasonably well. However, it is clear that the quality of the agreement depends on having a good description not only of the total flux, but of the input X-ray drive spectrum, in particular the sub-keV region where photo-ionization cross sections for the outer electrons are large. For L-shell spectra in the >1 keV region, crystal spectroscopy is a relatively robust diagnostic, and the calibration of X-ray crystals is commonplace. In the < 1 keV range we have a couple of choices, one being the use of diode arrays such as the DANTE system [24]. This can give absolute fluxes but likely with a low spectral resolution, depending on the number of diodes and the filter choices. On the other hand, the strong dependence of photoionization cross sections on photon energy near edges means that it may be appropriate to also field XUV grating spectrometers, to allow a more detailed understanding of the spectral shape at <100 eV.

The almost certain possibility of changing opacity in the lower photon energy region is of importance, and for that reason, time resolution measuring the lower energy photon spectrum is of importance. Thus far, experimentally, we have the temporal history of the L-shell flux and have assumed the same temporal shape for the lower photon flux. This is by no means certain and, for example, it is possible that lower photon energy flux will start earlier and a degree of pre-ionisation will occur before fluorescence is generated by the L-shell photons.

### 5.1. Experimental future work

Our work here shows that the time resolving capability of Cloudy is suitable for modelling laboratory plasmas where ionisation is occurring on nanosecond timescales. However, it is still desirable to create experiments where we can create intense fluxes with photo-ionization parameters $\xi$ ~10-100 erg-cms$^{-1}$ but on timescales closer to the quasi-steady state expected of astrophysical plasmas. We have hence planned future experiments where instead of double-sided irradiation with Ag/CH foils, as in figure 1, we will replace one foil with a hohlraum target with an Au inner coating. Careful control of intensity to suppress the 3-4 keV M-shell radiation [25] will create an intense broadband source, principally in the sub-keV regime, that will pass through an exit hole into the sample gas cell. By timing the beams irradiating the hohlraum to arrive earlier than those driving the Ag/CH foil we can generate a pre-ionised plasma that is probed by the K-shell fluorescence measurements only after significant evolution. Our initial calculations indicate that we can extend the characteristic time for the evolution of the average ionisation from the 0.2 – 2 ns in our previous experiments to 3 – 20 ns.

An option to also consider in future experiments is the use of time-gated instruments to record the K-shell fluorescence emission spectrum. This should allow a more detailed comparison with simulations than time-integrated data which will always contain signal from lower ion stages which are closer together in wavelength and thus harder to resolve than the emission from higher stages. Good spatial resolution for the detector surface would be necessary to maintain the high spectral resolution required to separate the K-$\alpha$ and K-$\beta$ lines from different ion stages. Resolutions of 50 $\mu$m with gating times down to 0.1 ns have been reported [26] and this corresponds to about 1 eV in spectral resolution for the spectrometer used in this work.

### 5.2. Cloudy at high densities

The goal of Cloudy is to yield reliable results for plasma densities ranging from the low-density limit to values so high that the system is in Local Thermodynamic Equilibrium (LTE) or Strict Thermodynamic Equilibrium (STE). This is a challenging task due to the uncertain physics of highly excited states. The 2017 Cloudy release paper [27] goes into some details. Our models of one and two electron systems are well behaved at all densities and level populations reach the proper thermodynamic limits at high densities, as shown in Fig 10 and 11 of that paper. Figs 17 and 18 of the 2013 release paper [28] show that the chemistry, ionization, and energy exchange go to the proper thermodynamic limits for a broad range of densities.

The 2025 Cloudy release [29] includes a major expansion in the treatment of excited states (Dehghanian+ 2025 in preparation) with the adoption of a large body of atomic data computed with the R-matrix suite of codes [30]. This improves the physical treatment of the highest levels that mediate the approach to statistical equilibrium.

However, substantial questions remain, with the theory of continuum lowering at high densities being the greatest uncertainty. Alimohamadi & Ferland [31] discuss continuum lowering and its effects on the partition function, and Section 3 of that paper shows that the three available theories for continuum lowering at high densities disagree by distressing amount. A proper theory of dense-plasma continuum lowering remains an unsolved grand challenge problem in physics.

Dielectronic recombination is often the dominant process for complex ions [32]. This occurs through highly excited and autoionizing states that are greatly affected by continuum lowering. Nigel Badnell and coworkers have created a theory for this suppression and provided numerical fits to density-dependent dielectronic recombination suppression factors [33], [34], [35]. These results are used by Cloudy to account for the high-density suppression of recombination.

The Cloudy team participated in two of the "NLTE#" series of meetings [36], [37], which compared predictions of codes designed for dense plasma laboratory experiments. Discussions at these workshops suggested that the leading cause for disagreement between predictions of the various codes was the treatment of continuum lowering upon dielectronic recombination. This remains an uncertainty.

## 6. Acknowledgements

This work was supported by the Leverhulme Trust via grant award RPG-2022-241 and the UKRI Science and Technology Facilities Council through grant ST/P000312/1. We would like to thank the UK Central Laser Facility staff who run the laser, target area and target preparation facilities for their contributions.

## 7. Data availability

The Cloudy models and experimental spectra used in this paper will be made available on request.

## 8. References

[1] Shapiro Stuart L. and Teukolsky Saul A., *Black Holes, White Dwarfs and Neutron Stars: The Physics of Compact Objects*. NewYork: John Wiley & Sons, 1983.

[2] Shakura N.I. and Sunyaev R.A., "Black holes in binary systems. Observational appearance.," *Astron. & Astrophys.*, vol. 24, p. 337355, 1973, doi: 1973A&A....24..337S.

[3] H. Meusinger and V. Weiss, "Ultraviolet variability of quasars: dependence on the accretion rate," *Astron Astrophys*, vol. 560, p. A104, Dec. 2013, doi: 10.1051/0004-6361/201322410.

[4] V. Connaughton *et al.*, "FERMI GBM observations of LIGO Gravitational-Wave event GW150914," *Astrophys J Lett*, vol. 826, no. 1, p. L6, Jul. 2016, doi: 10.3847/2041-8205/826/1/L6.

[5] J. E. Pringle, "Accretion Discs in Astrophysics," *Annu Rev Astron Astrophys*, vol. 19, no. 1, pp. 137–160, Sep. 1981, doi: 10.1146/annurev.aa.19.090181.001033.

[6] B. P. Abbott *et al.*, "GW151226: Observation of Gravitational Waves from a 22-Solar-Mass Binary Black Hole Coalescence," *Phys Rev Lett*, vol. 116, no. 24, p. 241103, Jun. 2016, doi: 10.1103/PhysRevLett.116.241103.

[7] T. Kallman *et al.*, "A census of x-ray gas in NGC 1068: results from 450 ks of CHANDRA high energy transmission grating observations," *Astrophys J*, vol. 780, no. 2, p. 121, Dec. 2013, doi: 10.1088/0004-637X/780/2/121.

[8] M. Chatzikos *et al.*, "THE 2023 RELEASE OF Cloudy," *Rev Mex Astron Astrofis*, vol. 59, no. 2, pp. 327–343, Oct. 2023, doi: 10.22201/IA.01851101P.2023.59.02.12.

[9] C. M. Gunasekera, P. A. M. van Hoof, M. Chatzikos, and G. J. Ferland, "The 23.01 Release of Cloudy," *Res Notes AAS*, vol. 7, no. 11, p. 246, Nov. 2023, doi: 10.3847/2515-5172/ad0e75.

[10] M. E. Foord *et al.*, "Charge-state distribution and doppler effect in an expanding photoionized plasma," *Phys Rev Lett*, vol. 93, no. 5, Jul. 2004, doi: 10.1103/PhysRevLett.93.055002.

[11] S. Fujioka *et al.*, "X-ray astronomy in the laboratory with a miniature compact object produced by laser-driven implosion," *Nat Phys*, vol. 5, no. 11, pp. 821–825, 2009, doi: 10.1038/nphys1402.

[12] G. P. Loisel *et al.*, "Benchmark Experiment for Photoionized Plasma Emission from Accretion-Powered X-Ray Sources," *Phys Rev Lett*, vol. 119, no. 7, Aug. 2017, doi: 10.1103/PhysRevLett.119.075001.

[13] R. C. Mancini *et al.*, "X-ray heating and electron temperature of laboratory photoionized plasmas," *Phys Rev E*, vol. 101, no. 5, May 2020, doi: 10.1103/PhysRevE.101.051201.

[14] D. C. Mayes *et al.*, "Observation of ionization trends in a laboratory photoionized plasma experiment at Z," *Phys Rev E*, vol. 104, no. 3, Sep. 2021, doi: 10.1103/PhysRevE.104.035202.

[15] S. White *et al.*, "Production of photoionized plasmas in the laboratory with x-ray line radiation," *Phys Rev E*, vol. 97, no. 6, Jun. 2018, doi: 10.1103/PhysRevE.97.063203.

[16] E. G. Hill and S. J. Rose, "Alternative methods of producing photoionised plasmas in the laboratory," *High Energy Density Phys*, vol. 7, no. 4, pp. 377–382, Dec. 2011, doi: 10.1016/j.hedp.2011.05.010.

[17] D. Riley *et al.*, "Generation of photoionized plasmas in the laboratory of relevance to accretion-powered x-ray sources using keV line radiation," *High Energy Density Phys*, vol. 51, Jun. 2024, doi: 10.1016/j.hedp.2024.101097.

[18] R. L. Singh *et al.*, "L-Shell X-Ray Conversion Yields for Laser-Irradiated Tin and Silver Foils," *Laser and Particle Beams*, vol. 2022, 2022, doi: 10.1155/2022/3234804.

[19] J. T. Larsen and S. M. Lane, "HYADES—A plasma hydrodynamics code for dense plasma studies," *J Quant Spectrosc Radiat Transf*, vol. 51, no. 1, pp. 179–186, 1994, doi: https://doi.org/10.1016/0022-4073(94)90078-7.

[20] W. J. Henney, S. J. Arthur, R. J. R. Williams, and G. J. Ferland, "Self-Consistent Dynamic Models of Steady Ionization Fronts. I. Weak-D and Weak-R Fronts," *Astrophys J*, vol. 621, no. 1, pp. 328–347, Mar. 2005, doi: 10.1086/427491.

[21] W. J. Henney, R. J. R. Williams, G. J. Ferland, G. Shaw, and C. R. O'Dell, "Merged Ionization/Dissociation Fronts in Planetary Nebulae," *Astrophys J*, vol. 671, no. 2, pp. L137–L140, Dec. 2007, doi: 10.1086/525023.

[22] S. J. Rose, "Calculations of the radiative opacity of laser-produced plasmas," *Journal of Physics B: Atomic, Molecular and Optical Physics*, vol. 25, no. 7, pp. 1667–1681, Apr. 1992, doi: 10.1088/0953-4075/25/7/034.

[23] H.-K. Chung, M. H. Chen, W. L. Morgan, Y. Ralchenko, and R. W. Lee, "FLYCHK: Generalized population kinetics and spectral model for rapid spectroscopic analysis for all elements," *High Energy Density Phys*, vol. 1, no. 1, pp. 3–12, 2005, doi: https://doi.org/10.1016/j.hedp.2005.07.001.

[24] E. L. Dewald *et al.*, "Dante soft x-ray power diagnostic for National Ignition Facility," *Review of Scientific Instruments*, vol. 75, no. 10, pp. 3759–3761, Oct. 2004, doi: 10.1063/1.1788872.

[25] D. R. Kania *et al.*, "Characterization of an x-ray-flux source for the production of high-energy-density plasmas," 1992.

[26] J. A. Oertel *et al.*, "Gated x-ray detector for the National Ignition Facility," *Review of Scientific Instruments*, vol. 77, no. 10, p. 10E308, Sep. 2006, doi: 10.1063/1.2227439.

[27] G. ~J. Ferland *et al.*, "The 2017 Release Cloudy," vol. 53, pp. 385–438, Oct. 2017, doi: 10.48550/arXiv.1705.10877.

[28] G. ~J. Ferland *et al.*, "The 2013 Release of Cloudy," vol. 49, pp. 137–163, Apr. 2013, doi: 10.48550/arXiv.1302.4485.

[29] C. M. Gunasekera *et al.*, "The 2025 Release of Cloudy," 2025. [Online]. Available: https://arxiv.org/abs/2508.01102

[30] G. Del Zanna, G. Liang, J. Mao, and N. R. Badnell, "UK APAP R-Matrix Electron-Impact Excitation Cross-Sections for Modelling Laboratory and Astrophysical Plasma," *Atoms*, vol. 13, no. 5, p. 44, May 2025, doi: 10.3390/atoms13050044.

[31] P. Alimohamadi and G. ~J. Ferland, "A Practical Guide to the Partition Function of Atoms and Ions," vol. 134, no. 1037, p. 73001, Jul. 2022, doi: 10.1088/1538-3873/ac7664.

[32] D. E. Osterbrock and G. J. Ferland, *Astrophysics of gaseous nebulae and active galactic nuclei*. 2006.

[33] D. Nikolić, T. ~W. Gorczyca, K. ~T. Korista, G. ~J. Ferland, and N. ~R. Badnell, "Suppression of Dielectronic Recombination due to Finite Density Effects," vol. 768, no. 1, p. 82, May 2013, doi: 10.1088/0004-637X/768/1/82.

[34] T. ~W. Gorczyca, D. Nikolić, K. ~T. Korista, N. ~R. Badnell, and G. ~J. Ferland, "Suppression of Dielectronic Recombination at Finite Densities," in *Journal of Physics Conference Series*, in Journal of Physics Conference Series, vol. 488. IOP, Apr. 2014, p. 62027. doi: 10.1088/1742-6596/488/6/062027.

[35] D. Nikolić *et al.*, "Suppression of Dielectronic Recombination Due to Finite Density Effects. II. Analytical Refinement and Application to Density-dependent Ionization Balances and AGN Broad-line Emission," vol. 237, no. 2, p. 41, Aug. 2018, doi: 10.3847/1538-4365/aad3c5.

[36] H.-K. Chung, C. Bowen, C. ~J. Fontes, S. ~B. Hansen, and Yu. Ralchenko, "Comparison and analysis of collisional-radiative models at the NLTE-7 workshop," *High Energy Density Phys*, vol. 9, no. 4, pp. 645–652, Dec. 2013, doi: 10.1016/j.hedp.2013.06.001.

[37] R. Piron *et al.*, "Review of the 9th NLTE code comparison workshop," *High Energy Density Phys*, vol. 23, pp. 38–47, Jun. 2017, doi: 10.1016/j.hedp.2017.02.009.